\documentstyle[pra,aps,epsfig,array,twocolumn]{revtex}
\newcommand{\upp}[1]{^{\rm \scriptscriptstyle #1}}
\newcommand{\dnn}[1]{_{\rm \scriptscriptstyle #1}}
\newcommand{\ket}[1]{| \, #1 \rangle}
\newcommand{\bra}[1]{ \langle #1 \,  |}
\newcommand{\proj}[1]{\ket{#1}\bra{#1}}
\newcommand{\pket}[2]{ | \, #1 \rangle \otimes | \, #2  \rangle  }

\newcommand{\ppket}[3]{ | \, #1 \rangle \otimes  | \, #2 \rangle
                        \otimes  | \, #3 \rangle }

\newcommand{\abs}[1]{ | \, #1 \,  |}

\newcommand{\asred}{\preceq}
\newcommand{\asredchem}{\leadsto}
\newcommand{\asequ}{\approx}
\newcommand{\exred}{\le}

\newcommand{\exequchem}{\rightleftharpoons}

\newcommand{\espan}{{\cal S}}
\newcommand{\tr}[2]{{\rm tr}_{\rm \scriptscriptstyle #1}(#2)}
\newcommand{\identity}{\mbox{\boldmath{1} \hspace{-0.33cm} \boldmath{1} }}

\newcommand{\beq}{\begin{equation}}
\newcommand{\eeq}{\end{equation}}
\newtheorem{lem}{Lemma}
\newtheorem{theo}{Theorem}

\newtheorem{cor}{Corollary}

\title{Exact and Asymptotic Measures of Multipartite Pure State Entanglement}
\author{Charles H. Bennett$^1$,
Sandu Popescu$^2$, Daniel Rohrlich$^3$,
John A. Smolin$^1$, and Ashish V. Thapliyal$^4$}
\address{$^1$IBM Research Division, Yorktown Heights, NY 10598, USA ---
{\tt bennetc, smolin@watson.ibm.com}}
\address{$^2$Isaac Newton Institute, Cambridge University,
Cambridge, UK and BRIMS, Hewlett-Packard Labs., Stoke Gifford, Bristol BS12 6QZ, UK --- {\tt sp230@newton.cam.ac.uk} }
\address{$^3$School of Physics and Astronomy,Tel Aviv University, Tel Aviv, 
Israel --- {\tt rohrlich@post.tau.ac.il}}
\address{$^4$ Dept. of Physics, Univ. of
California, Santa Barbara, CA 93106, USA, {\tt ash@physics.ucsb.edu}}

\date{\today}
\begin{document}
\twocolumn[\hsize\textwidth\columnwidth\hsize\csname
@twocolumnfalse\endcsname
\maketitle

\begin{abstract} Hoping to simplify the classification of pure
entangled states of multi ($m$)-partite quantum systems, we study
exactly and asymptotically (in $n$) reversible transformations among
$n^{\mathrm th}$ tensor powers of such states (ie $n$ copies of the
state shared among the same $m$ parties) under local quantum operations
and classical communication (LOCC). For exact transformations, we show
that two states whose marginal one-party entropies agree are either
locally-unitarily (LU) equivalent or else LOCC-incomparable. In
particular we show that two tripartite Greenberger-Horne-Zeilinger (GHZ)
states are LOCC-incomparable to three bipartite Einstein-Podolsky-Rosen
(EPR) states symmetrically shared among the three parties. Asymptotic
transformations yield a simpler classification than exact
transformations; for example, they allow all pure bipartite states to be
characterized by a single parameter---their partial entropy---which may
be interpreted as the number of EPR pairs asymptotically
interconvertible to the state in question by LOCC transformations. We
show that $m$-partite pure states having an $m$-way Schmidt
decomposition are similarly parameterizable, with the partial entropy
across any nontrivial partition representing the number of standard
``Cat'' states $\ket{0^{\otimes m}}+\ket{1^{\otimes m}}$ asymptotically
interconvertible to the state in question. For general $m$-partite
states, partial entropies across different partitions need not be equal,
and since partial entropies are conserved by asymptotically reversible
LOCC operations, a multicomponent entanglement measure is needed, with
each scalar component representing a different kind of entanglement, not
asymptotically interconvertible to the other kinds. In particular we
show that the m=4 Cat state is not isentropic to, and therefore not
asymptotically interconvertible to, any combination of bipartite and
tripartite states shared among the four parties. Thus, although the m=4
cat state can be prepared from bipartite EPR states, the preparation
process is necessarily irreversible, and remains so even asymptotically.
For each number of parties $m$ we define a minimal reversible
entanglement generating set (MREGS) as a set of states of minimal
cardinality sufficient to generate all $m$-partite pure states by
asymptotically reversible LOCC transformations. Partial entropy
arguments provide lower bounds on the size of the MREGS, but for $m>2$
we know no upper bounds. We briefly consider several generalizations of
LOCC transformations, including transformations with some probability of
failure, transformations with the catalytic assistance of states other
than the states we are trying to transform, and asymptotic LOCC
transformations supplemented by a negligible ($o(n)$) amount of quantum
communication. 
\end{abstract} 
]


\section{Introduction}
Entanglement, first noted by  Einstein-Podolsky-Rosen (EPR)
\cite{ein:pod:ros:35} and
Schr$\ddot{\rm o}$dinger \cite{sch:35}, is an essential feature of
quantum mechanics. Entangled two-particle states, by their
experimentally verified violations of Bell inequalities, have played
an important role in establishing widespread confidence in the correctness of
quantum mechanics.  Three-particle entangled states, though more difficult to
produce experimentally, provide even stronger tests of quantum nonlocality.

The canonical two-particle entangled state is the Einstein-Podolsky-Rosen-Bohm
(EPR) pair,
\beq
\ket{00}+\ket{11}.
\eeq
(We omit normalization factors when it will cause no confusion).
The canonical tripartite entangled state is the
Greenberger-Horne-Zeilinger-Mermin (GHZ) state
\beq
\ket{000}+\ket{111},
\eeq
while the corresponding $m$-partite state
\beq
\ket{0^{\otimes m}}+\ket{1^{\otimes{m}}}.
\eeq
is called an $m$-particle Cat ( $m$-Cat) state,
in honor of Schr\"{o}dinger's cat.

More recently it has been realized that entanglement is a useful
resource for various kinds of quantum information processing,
including quantum state teleportation~\cite{ben:bra:cre:joz:per:woo:93},
cryptographic key distribution~\cite{ben:bra:84}, classical
communication over quantum channels \cite{ben:wie:92,ben:fuc:smo:97,ben:sho:smo:tha:99},
quantum error correction~\cite{got:97}, quantum computational speedups~\cite{deu:85},
and distributed computation~\cite{gro:97,cle:bur:97}.
In view of its central role~\cite{rev:phywor:98} in quantum
information theory, it is important to have a qualitative
and quantitative theory of entanglement.

Entanglement only has meaning in the context of a multipartite quantum
system, whose Hilbert space can be viewed as a product of two
or more tensor factors corresponding physically to subsystems of the system.
We often think of subsystems as belonging to different observers, e.g. Alice
has subsystem A, Bob has subsystem B and so on.

Mathematically, an {\em unentangled} or {\em separable} state is a mixture
of product states; operationally it is a state that can be made from a pure 
product state by local operations and classical communication (LOCC). Here local
operations include unitary transformations, additions of ancillas
(ie enlarging the Hilbert space), measurements, and throwing
away parts of the system, each  performed by one party on his/her
subsystem. Mathematically, we represent LOCC by a {multilocal}
superoperator, i.e. a completely positive linear map that does not
increase the trace, and can be implemented locally with  classical
coordination  among the parties.\footnote{ General quantum dynamics
can be represented mathematically by completely positive linear
maps that do not increase trace~\cite{hel:kra:69:70:83,cav:98}.
Such a map say ${\cal L}$ can be written as  ${\cal
L}(\rho) = \sum_i L_i\rho L_i^{\dagger}$, where
$\sum_iL_i^{\dagger} L_i\le\identity$. The equality holds for
trace-preserving superoperators which correspond physically to
non-selective dynamics, e.g. a measurement followed by forgetting
which outcome was produced. In general the superoperators may be
trace decreasing and correspond to selective operations, e.g. a
measurement followed by throwing away some outcomes. If
${\cal L}$ is a multilocally implementable superoperator, it must
be a separable superoperator, i.e. a completely positive
trace-preserving map of the form shown above, where the $L_i$'s  are
products of local operators --
$L_i=L\upp{A}_i\otimes L\upp{B}_i\ldots$ \ . Note that not all
separable superoperators are multilocally
implementable~\cite{ben:div:fuc:mor:rai:sho:smo:woo:98,rai:98}. }
Classical communication between parties allows local actions by
one party to be conditioned on the outcomes of earlier
measurements performed by other parties.  This allows, among other
things, the creation of
 mixed states that are classically correlated but not entangled.

Mathematically speaking, a pure state $\ket{\Psi\upp{ABC...}}$
is separable if and only if it can be expressed
as a tensor product of states belonging to different parties:
\begin{equation}
\ket{\Psi\upp{ABC...}} = \ppket{\alpha\upp{A}}{\beta\upp{B}}{\gamma\upp{C}}
\otimes ... \enspace.
\end{equation}
A mixed state $\rho\upp{ABC...}$ is separable if and only if it can be
expressed as a mixture of separable pure states:
\begin{equation}
\rho\upp{ABC...} =
\sum_i p_i \proj{\alpha_i\upp{A}} \otimes \proj{\beta_i\upp{B}}
\otimes \proj{\gamma_i\upp{C}}
\otimes ... \enspace,
\end{equation}
where the probabilities $p_i\ge0$ and $\sum_i p_i = 1.$

States that are not separable are said to be
{\em entangled} or {\em inseparable}.

Besides the gross distinction between entangled and unentangled states,
various inequivalent {\em kinds\/} of entanglement can be distinguished,
in recognition of the fact that not all entangled states can be
interconverted by local operations and classical communication. For
example, bipartite entangled states are further subdivided into {\em
distillable\/} and {\em bound\/} entangled states, the former being states
which are pure or from which some pure entanglement can be produced
by LOCC, while the latter are mixed states which, though inseparable, have zero
distillable entanglement.

Within a class of states having the same kind of entanglement (eg
bipartite pure states) one can seek a scalar measure of
entanglement.  Five natural desiderata for such a measure~(cf.
\cite{ben:ber:pop:sch:96,bbpssw96,pop:roh:97,ved:ple:rip:kni:97,vid:99}) are:
\begin{itemize}
\item It should be zero for separable states.
\item It should be invariant under local unitary transformations.
\item Its expectation should not increase under LOCC.
\item It should be additive for tensor products of independent states, shared
among the same set of observers (thus if $\Psi\upp{AB}$ and $\Phi\upp{AB}$ are
are bipartite states shared between Alice and Bob, and $E$ is an entanglement measure,
$E(\Psi\upp{AB}\otimes\Phi\upp{AB})$ should equal $E(\Psi\upp{AB})+E(\Phi\upp{AB})$).
\item It should be stable \cite{kit:98} with respect to transfer of a subsystem from
one party to another, so that in any tripartite state $\Psi\upp{ABC}$, the bipartite
entanglement of $AB$ with $C$ should differ from that of $A$ with $BC$ by at most
the entropy of subsystem $B$.
\end{itemize}

For bipartite pure states it has been shown \cite{ben:ber:pop:sch:96,pop:roh:97,vid:99} that
asymptotically there is only one kind of entanglement and partial
entropy is a good entanglement measure ($E$) for it. It is equal,
both to the state's
{\em entanglement of formation} (the number of EPR pairs asymptotically
required to prepare the state by LOCC), and the state's {\em distillable entanglement}
(the number of EPR pairs asymptotically preparable from the state by LOCC).
Here partial entropy is the Von Neumann entropy
$S(\rho)=\tr{}{\rho\log_2\rho}$
of the reduced density matrix obtained by tracing out either of
the two parties.

In section \ref{sec:redequ} to follow, we define
exact and asymptotic reducibilities and equivalences under LOCC alone, 
and with the help of ``catalysis'', or asymptotically negligible
amounts of quantum communication.  
In section \ref{sec:mulent} we use these concepts to
develop a framework for quantifying tripartite and multipartite
pure-state entanglement, in terms of a canonical set of states
which we call a minimal reversible entanglement generating set (MREGS).
This framework leads to an additive, multicomponent entanglement
measure, based on asymptotically reversible LOCC transformations
among tensor powers of such states, and having a number of scalar
components equal to the number of states in the MREGS, in other
words the number of asymptotically inequivalent kinds of
entanglement. 

For general $m$-partite states, partial entropy arguments 
give lower bounds on the number of entanglement components 
as a function of $m$, and allow us to show that some states, 
e.g. the $m=4$ Cat state, are not exactly, nor even asymptotically,
interconvertible into any combination of EPR pairs shared among
the parties.  

On the other hand, we show that the
subclass of multipartite pure states having an $m$-way Schmidt
decomposition is describable by a single parameter, its partial
entropy representing the number of standard ``Cat'' states
$\ket{0^{\otimes m}}+\ket{1^{\otimes m}}$ asymptotically
interconvertible to the state in question.  Section
\ref{sec:EPRGHZ} treats tripartite pure state entanglement,
showing in particular that, using exact LOCC transformations, two
GHZ states can neither be prepared from nor used to prepare the
isentropic combination of three EPR pairs shared symmetrically
among the three parties.

\section{Reducibilities, Equivalences and Local Entropies}
\label{sec:redequ}

Reducibility formalizes the notion of a transformation of one 
state to another being possible under certain conditions, while equivalence 
formalizes the notion of this transformation being reversible---possible 
in both directions.  
While studying entanglement it is useful to discuss state transformation
under LOCC. This is because  a good entanglement measure should not
increase under LOCC.  So, if
two states are equivalent under LOCC operations, they will have
the same entanglement. This is the key idea we will use in section
\ref{sec:mulent} to quantify entanglement.

We start by first looking at partial entropies. Partial entropies have
the nice property that for pure states their average does not
increase under LOCC.

Suppose the $m$ parties holding a pure state $\Psi$ are
numbered 1,2,...$m$. Let $X$ denote a nontrivial subset of the
parties and let $\bar{X}$ be the set of remaining parties.
Then the reduced density matrix of subset $X$ of the
parties is defined as \beq \rho\dnn{\mathit X}(\Psi)=\tr{ \bar{X}}{\proj{\Psi}}.
\eeq The {\em partial entropy\/} of subset $X$ is the von
Neumann entropy 
\beq S_X(\Psi) =
-\tr{}{\rho\dnn{\mathit X}(\Psi)\log_2\rho\dnn{\mathit X}(\Psi))}. 
\eeq
When $X=\{\ell\}$ consists of a single party, $\rho_{\{\ell\}}$ is called the 
{\em marginal density matrix\/} of party $\ell$ and $S(\rho_{\{\ell\}})$ its
the {\em marginal entropy\/} of party $\ell$.
Two states are said to be {\em isentropic} if for each subset $X$ of the parties
$S_X(\Psi)=S_X(\Phi)$. Two states $\Psi$ and $\Phi$ are said to be 
{\em marginally isentropic} if $S_{\{\ell\}}(\Psi)=S_{\{\ell\}}(\Phi)$ 
for each party $\ell$.

Now we are ready to show that for any subset $X$ of parties, the partial
entropy $S_X$ is nonincreasing under LOCC.  We state this as a lemma,
\begin{lem}
\label{lem:noninc}
{\rm {\bf :} If a multipartite system is
initially in a pure state $\Psi$, and is subjected to a sequence of LOCC
operations resulting in a set of final pure states $\Phi_i$ with
probabilities $p_i$, then
for any subset $X$ of the parties }
\beq
S_X(\Psi) \geq \sum_i p_i \;S_X(\Phi_i)
\label{LEnoninc}
\eeq
\end{lem}

\noindent{\bf Proof:}
The result follows from the fact  that average bipartite
entanglement (partial entropy) of bipartite pure states cannot
increase under LOCC cf. \cite{ben:div:smo:woo:96}.
\medskip

\subsection{Reducibilities and equivalences: exact and stochastic}

We start with LOCC state transformation involving single copies of
states. If the state transformation is exact we say it is an
exact reducibility. If the state transformation suceeds some of
the time we say it is stochastic, and if the state transformation
needs the presence of another state, which is is recovered after the
protocol, it is called catalytic reducibility. In this section
we define these more precisely. We start with exact reducibility.

We say a state $\Phi$ is {\em exactly reducible}  to a state
$\Psi$ (written $\Phi\exred\dnn{LOCC}\Psi$ or just $\Phi\exred\Psi$)
by local operations and classical communication if and only if
\begin{equation}
\exists_{\cal L}\;\;\;
\Phi = {\cal L}(\Psi)\enspace,
\end{equation}
where ${\cal L}$ is a multilocally implementable trace preserving
superoperator. Alternatively we may say that the LOCC protocol ${\cal P_L}$
corresponding to the superoperator ${\cal L}$ transforms $\Psi$ to $\Phi$
exactly.

Intuitively this means that the state transformation from $\Psi$ to $\Phi$
can be done by LOCC with probability one. (Where it
will cause no confusion, for pure states we use a plain Greek letter
such as $\Psi$ to
represent both the vector $\ket{\Psi}$ and the projector $\proj{\Psi}$.)

The relation of exact LOCC reducibility for bipartite pure states
has been studied in \cite{lo:pop:97} and{\cite{nie:98}, which give necessary and
sufficient conditions for it in terms of majorization of the eigenvalues
of the reduced density matrix.
Nielsen (\cite{nie:98}) uses notation reminiscent of a chemical
reaction: where we say $\Phi\exred\Psi$, he says
$\Psi\rightarrow\Phi$. Both notations mean that given one
copy of $\Psi$, we can with certainty, by local operations and classical
communication, make one copy of $\Phi$.

Chemical reactions often involve catalysts, molecules which facilitate a
reaction without being used up, so it is natural to look for analogous
quantum state transformations. Jonathan and Plenio have recently found 
an example of successful catalysis for bipartite states, where a catalyst 
allows a transformation to be performed with certainty which could only 
be done with some chance of failure in the absence of the the catalyst \cite{jon:ple:99}.

We say that  $\Phi$ is {\em catalytically
reducible\/} ($\exred\dnn{LOCCc}$) to $\Psi$ if and only if there exists a
state  $\Upsilon$ such that
\beq
\Phi\otimes\Upsilon \exred\dnn{LOCC} \Psi\otimes\Upsilon.
\label{catred}\eeq

An interesting fact about catalysis is that, because the catalyst is not consumed, 
one copy of it is sufficient to transform arbitrarily many copies of $\Psi$ into $\Phi$: 
\beq
\forall_n \;\Phi \Upsilon \exred\dnn{LOCC}\Psi \Upsilon \Rightarrow  
\Phi^n \Upsilon \exred\dnn{LOCC} \Psi^n \Upsilon.\label{stretchcatyl}
\eeq

Another important form of state transformation involves probabilistic
outcomes, where the procedure for the reducibility may fail some of the
time as in ``entanglement gambling''~\cite{ben:ber:pop:sch:96}.
We capture this idea in stochastic reducibility:

We say a state $\Phi$ is {\em stochastically  reducible}  to a
state $\Psi$ 
under LOCC with yield $p$ if and only if
\begin{equation}
\exists_{{\cal L}}\;\;\;
\Phi = \frac{{\cal L}(\Psi)}{\tr{}  {{\cal L} (\Psi)}} \enspace,
\end{equation}
where ${\cal L}$ is a multilocally implementable superoperator
such that $\tr{}{{\cal L}(\Psi)} = p$. 

This means that a copy of $\Phi$ may be obtained from a copy 
of $\Psi$ with probability $p$ by LOCC operations.  Exact reducibility
corresponds to the case $ p\!=\!1$.

For any reducibility, one may define corresponding notions of
equivalence and incomparability.

Two states $\Phi$ and $\Psi$are said to be {\em exactly equivalent}
($\equiv\dnn{LOCC}$ or simply $\equiv$ ) if $\Phi \exred \Psi$
and $\Psi\exred \Phi$.  
This means that the two states
are exactly interconvertible by classically coordinated local
operations. In chemical notation this would be
$\Psi\exequchem\Phi$. Conversely, states $\Phi$ and $\Psi$ are said 
to be {\em exacty incomparable\/} 
if neither is exactly reducible to the other.

Catalytic and stochastic equivalence and incomparability may be defined
analogously.~\footnote{Very recently D\"{u}rr, Vidal, and Cirac (LANL
eprint quant-ph/0005115) have found a tripartite pure state of 3 qubits
which is stochastically incomparable with the GHZ state. They also show
that if two pure states are chosen randomly in the tensor product
Hilbert space of four or more parties, then, with probability one, they
are stochastically incomparable: neither state can be produced from the
other by LOCC with any chance of success.}

In passing we note that many other reducibilities (and their
corresponding equivalences) can be considered, e.g. reducibilities via
local unitary operations~\cite{lin:pop:97}
$\exred\dnn{LU}$, stochastic reducibility with catalysis, and
reducibilities without communication or with one-way communication~\cite{ben:97}.

Physically, reducibility via local unitary operations and that via 
local unitary operations along with a change in the local support 
(corresponding to the
increase or decrease in the local Hilbert space dimensions) are the same
because we could think of the extra dimensions as being present from the
start and extend the local unitary operation to the larger space. Thus, 
from now on when we say local unitary operations, we mean local unitary 
operations along with a possible change in the local support, i.e., isometric 
transformations.\footnote{ Unitary operations are characterized by 
$U^\dagger U=\identity=UU^{\dagger}$. However, if we are want general
transformations that preserve the norm of vectors, all we need is 
$U^{\dagger}U=\identity$, where the $U$'s could be rectangular matrices.
Such $U$ are called isometric\cite{weh:78}.}

We now look at some conditions for two states to be exactly
equivalent. By Lemma \ref{lem:noninc} it is clear that if two states are
equivalent they must be isentropic, but not all isentropic states
are equivalent.

We are now in a position to demonstrate some important facts about exact LOCC
reducibility\footnote{These results strengthen Guifre Vidal's result 
\cite{vid:99} that LU equivalence $\Leftrightarrow$ 
LOCC equivalence for bipartite pure states, and Julia 
Kempe's result \cite{kempe} that if two multipartite pure 
states have isospectral marginal density matrices, then 
they are either LU-equivalent or LOCC-incomparable.}.
\begin{theo}{\rm {\bf :}
\label{theo:isenLU}
If $\Psi$ and $\Phi$ are two marginally isentropic pure
states, then they are either locally unitarily (LU)-equivalent or
else LOCC-incomparable.  
}\end{theo}
\begin{cor}
\label{cor:LULOCC}
{\rm {\bf :}
Two states are LOCC equivalent if and only if they are
LU equivalent.
\beq
\forall_{\Psi,\Phi} \;\Psi\equiv\dnn{LOCC}\!\Phi\;\Leftrightarrow\;\;\Psi\equiv\dnn{LU}\!\Phi.
\eeq }
\end{cor}
\begin{cor}
\label{cor:MARGINCOMP}
{\rm {\bf :} States that are marginally but not fully isentropic are
necessarily LOCC-incomparable. }
\end{cor}

\noindent{\bf Proof:}
To prove this it suffices to show that for marginally isentropic states
$\Psi$ and $\Phi$, if $\Phi
\exred \Psi$ then they must be local unitarily equivalent.
In light of the
non-increase of partial entropy under \mbox{LOCC}
(cf. lemma \ref{lem:noninc}) and the fact that these two
states are marginally isentropic, a LOCC protocol that
converts one state to the other must conserve the marginal entropies at
each step. Suppose the LOCC protocol ${\cal P}$ transforms  $\Psi$ to
$\Phi$ exactly. In general such a protocol consists of
a sequence of local transformations each done by one party followed
by communication of (some of) the information gained to other
parties.
Without loss of generality assume that Alice performs the first 
operation of such a protocol converting $\Psi$ to $\Phi$,
which gives the resulting ensemble ${\cal E}=\{p_i,\psi_i\}$.
Since Alice's operation cannot change the density matrix $\rho\upp{BC...}$
seen by  the remaining parties,
\beq
\rho\upp{BC...}=\sum_i p_i \tr{A}{\proj{\psi_i}} \enspace.
\eeq
As argued earlier, the average entropy must not change, i.e.
\beq
S\dnn{BC...}(\Psi)=S\dnn{A}(\Psi)=\sum_i p_i S\dnn{BC...}(\psi_i).
\eeq
By the strict concavity of the von Neumann entropy \cite{weh:78}
each of the resultant states $\psi_i$ must have the same reduced density 
matrix, from the viewpoint of all the other parties besides Alice, as 
the original state $\Psi$ did:
\beq
\forall_i \tr{A}{\proj{\psi_i}} = \tr{A}{\proj{\Psi}}. 
\eeq
Therefore the states $\psi_i$ must be related by isometries
acting on Alice's Hilbert space alone: 
\beq
\ket{\psi_i}=U_i\upp{A}\otimes I\upp{BC...} \ket{\Psi}. 
\eeq where
$U_i\upp{A}$ are unitary transformations acting on Alice's
Hilbert space, which  may have more dimensions than the
support of $\ket{\Psi}$ in Alice's space (this would
correspond to Alice having unilaterally chosen to enlarge 
her Hilbert space, which she is always free to do). Thus Alice's
measurement process, which appears on its face to be a stochastic
process not entirely under her control, could in fact 
be faithfully simulated by having her simply toss a coin to choose
a ``measurement result'' $i$ with probability $p_i$, then
perform the deterministic operation $U_i$ on her portion of the
joint state, and then finally report the result 
$i$ to all the other parties.  In  the
next step of the protocol, another party  performs similar
operations and sends classical information as to which unitary it
performed and so on for each step. Thus the entire
protocol consists of local unitary transformations, enlargement of
Hilbert space and classical communication, maintaining at each
step the overall state to be pure. The protocol ends when the
state $\Phi$ has been obtained. Since this is an exact
reducibility of one pure state to another, for each possible
sequence of local unitaries, the result must be $\Phi$. Thus we
can define a new protocol ${\cal P'}$ that consists of choosing
just one such sequence of local unitaries and it will take $\Psi$
to $\Phi$, showing that the two states are local unitarily
equivalent.  The first corollary follows from the fact that if two states are
LOCC equivalent, they must be isentropic and therefore marginally isentropic.
The second follows from the fact if that the two states were LU equivalent,
they would be fully isentropic, not merely marginally so.

\subsection{Asymptotic reducibilities and equivalences, and their
relation to partial entropies}

Before we discuss asymptotic reducibilities and equivalences, let
us define a quantitative measure of similarity of two states.
One such measure, the {\em fidelity}  \cite{uhl:76}\cite{joz:94}
of a mixed state $\rho$ relative to a pure state $\psi$, is given by
$F(\rho,\psi)=\bra{\psi}{\rho}\ket{\psi}$. It is the probability
that $\rho$ will pass a test for being $\psi$, conducted by
an observer who knows the state $\psi$. For mixed states $\rho$ and $\sigma$
it is given by the more symmetric expression
$F(\rho,\sigma)=({\rm tr}(\sqrt{\sigma}{\rho}\sqrt{\sigma})^{\frac{1}{2}})^2$.

Exact reducibility is too weak a reducibility to give a {\em simple}
classification of entanglement---even for bipartite pure states, there
are infinitely many incomparable $\exred\dnn{LU}$ equivalence classes,
which would lead to infinitely many distinct kinds of bipartite
entanglement. Linden and Popescu~\cite{lin:pop:97} have explored the
orbits of multipartite states under local unitary operations, and shown
that the number of LU invariants increases exponentially with the number
of parties and with the number of qubits possessed by each party.

One natural way to strengthen the notion of reducibility is to make
it asymptotic.  We first consider  ``asymptotic LOCC
reducibility''~\cite{ben:ber:pop:sch:96,ben:97} which expresses
the ability to convert $n$ copies of one pure state into a good
approximation of $n$ copies of another, in the limit of large $n$.
A possibly stronger  reducibility, which we will call
``asymptotic LOCCq reducibility,'' expresses the ability to do
the state transformation with the help of a limited ($o(n)$) amount of quantum
communication, in addition to the unlimited classical
communication and local operations allowed in ordinary LOCC
reducibility. Another natural way of strengthening asymptotic
reducibility is to allow catalysis; defining ``catalytic
asymptotic LOCC reducibility'' (LOCCc) in direct
analogy with the exact case.  We show that asymptotic LOCCc 
reducibility is at least as strong as (ie can simulate) LOCCq reducibility. 

Ordinary asymptotic LOCC reducibility is enough to simplify the classification of
all bipartite pure states and some classes of $m$-partite states, so that,
for any given $m$, a finite repertoire of standard states (EPR, GHZ, etc),
which we will later call a minimal reversible entanglement generating set
or MREGS, can be combined to prepare any member of class in an asymptotically
 reversible fashion, regardless of the size of the Hilbert spaces of the
parties.  Whether this classification can be extended to cover
general $m$-partite states for $m>2$ while maintaining a finite
repertoire size is an open question.

Let us start by defining ordinary asymptotic LOCC reducibility.

State $\Phi$ is {\em asymptotically reducible} ($\asred\dnn{LOCC}$ or
simply $\asred$) to
state $\Psi$ by local operations and classical communication if and only if
\begin{eqnarray}
\forall_{\delta>0,\epsilon>0} \;\exists_{n,n',{\cal L}} \;\;
\abs{(n/n')-1}<\delta \;\;{\rm and}& & \nonumber \\
F({\cal L}(\Psi^{\otimes n'}),\Phi^{\otimes n})
\ge 1-\epsilon \enspace & &. \label{defasred}
\end{eqnarray}
Here ${\cal L}$ is a multi-locally implementable superoperator
that converts $n'$ copies of $\Psi$ into a high fidelity approximation
to $n$ copies of $\Phi$.
In chemical notation we can write this as $\Psi \asredchem \Phi$.

A natural extension of asymptotic LOCC reducibility occurs if we
allow catalysis. Thus we define asymptotic LOCCc reducibility as:

\label{dfn:asredLOCCc}
 We say $\Phi$ is {\em asymptotically LOCCc reducible}
($\asred\dnn{LOCCc}$) to $\Psi$ if there exists some state $\Upsilon$
such that
\begin{equation}
\Phi \Upsilon \asred \Psi \Upsilon,\label{defasLOCCc}
\end{equation}
where we say the state $\Upsilon$ is a catalyst for this reducibility.
As with exact catalysis (eq.~\ref{stretchcatyl}), asymptotic catalysis
allows an arbitrarily large ratio of reactant to catalyst:
\begin{equation}
\Phi \Upsilon \asred \Psi \Upsilon
\Rightarrow \forall_n \Phi^n \Upsilon \asred \Psi^n \Upsilon.
\end{equation}

Another way of extending asymptotic LOCC reducibility is to allow a 
sublinear amount of quantum communication during the transformation
process.

State
$\Phi$ is said to be {\em asymptotically LOCCq reducible ($\asred\dnn{LOCCq}$)} to
state $\Psi$ iff
\begin{eqnarray}
\forall_{\delta>0,\epsilon>0} \;\exists_{n,k,{\cal L}} \;\;
(k/n)<\delta \;\;{\rm and}& & \nonumber \\
F({\cal L}(\Gamma^{\otimes k}\otimes\Psi^{\otimes n})
,\Phi^{\otimes n})  \ge 1-\epsilon \enspace & &, \label{defasLOCCq}
\end{eqnarray}
where $\Gamma$ denotes the $m$-Cat state $\ket{0^{\otimes m}}+\ket{1^{\otimes m}}$.

The $m$-Cat states used here are a convenient way of allowing a
sublinear amount $o(n)$ of quantum communication, since  they can be
used as described in section \ref{sec:EPRGHZ} to generate EPR pairs
between any two parties which in turn can be
used to teleport quantum data between the parties. The
$o(n)$ quantum communication allows the definition to be simpler
in one respect: a single tensor power $n$ can be used for the
input state $\Psi$ and output state $\Phi$, rather than the
separate powers $n$ and $n'$ used in the definition of ordinary
asymptotic LOCC reducibility without quantum communication,
because any $o(n)$ shortfall in number of copies of the output
state can be made up by using the Cat states to synthesize the
extra output states de novo. This definition is more natural than
that for ordinary asymptotic LOCC reducibility in that the input
and output states are allowed to differ in any way that can be
repaired by an $o(n)$ expenditure of quantum communication, rather
than only in the specific way of being $n$ versus $n'$ copies of
the desired state where $n-n'$ is $o(n)$.

Clearly $\asred\dnn{LOCC}$ implies $\asred\dnn{LOCCq}$ and $\asred\dnn{LOCCc}$,
because ordinary asymptotic reducibility is a special case of the
two other kinds of asymptotic reducibility. 
We can also show that asymptotic LOCCq reducibility implies
asymptotic LOCCc reducibility, because any $\asred\dnn{LOCCq}$ protocol
can be simulated by a $\asred\dnn{LOCCc}$ protocol with the $m$-Cat
state $\Gamma$ as catalyst, only a sublinear (and therefore asymptotically
negligible) amount of which is consumed.  In more detail, if
$\Phi\asred\dnn{LOCCq}\Psi$, then from eq.~\ref{defasLOCCq}, for each
$\epsilon$ and $\delta$, there exist $n$ and $k$ such that
$\Psi^{\otimes n}$ can be converted to a $1\!-\!\epsilon$ faithful
approximation to $\Phi^{\otimes n}$ with the help of $k<n\delta$ Cat
states' worth of quantum communication. This implies that $n$ copies of
$\Psi$ and $k$ copies of $\Gamma$ can be converted into a 
$1\!-\!\epsilon$ faithful approximation
to $n$ copies of $\Phi$ without any quantum communication. By supplying
$n\!-\!k$ extra, nonparticipatory copies of $\Gamma$, which are present
both before and after the transformation, and discarding $k$ of the
copies of $\Phi$ which the transformation has produced (even if the
copies are entangled, this cannot decrease the fidelity), we get that a
$1\!-\!\epsilon$~faithful approximation to
$(\Phi\otimes\Gamma)^{\otimes (n-k)}$ can be prepared from
$(\Psi\otimes\Gamma)^{\otimes n}$. This satisfies the conditions
(eq.~\ref{defasred}) for asymptotic reducibility
\beq
\Phi\otimes\Gamma \asred\dnn{LOCC} \Psi\otimes\Gamma,
\eeq
or, invoking the definition (eq.~\ref{defasLOCCc}) of asymptotic 
catalytic reducibility,
\beq
\Phi\asred\dnn{LOCCc} \Psi,
\eeq
which was to be demonstrated.  While the converse (i.e. that asymptotic catalytic
reducibility can be simulated by LOCCq transformations) seems plausible,
we have not been able to prove it except in special cases.

Asymptotic reducibilities and equivalences can have non-integer
yields.  This can be expressed using tensor exponents that take on
any nonnegative real value, so that $\Phi^{\otimes x} \asred \Psi^{\otimes y}$
denotes
\begin{eqnarray}
\forall_{\delta>0} \;\exists_{n,n',} \;\;
\abs{(n/n')-x/y}<\delta \;\;{\rm and}& & \nonumber \\
F({\cal L}(\Psi^{\otimes n'})
,\Phi^{\otimes n})  \ge 1-\epsilon \enspace & &.
\end{eqnarray}
In this case we say $x/y$ is the asymptotic efficiency or yield with
which $\Phi$ can be obtained from $\Psi$.
In chemical notation this could be expressed by
\mbox{$\Psi \asredchem  \frac{x}{y}\Phi$}, keeping in mind that the
coefficient $x$ represents an asymptotic yield or number of copies of
the state $\Phi$, not a scalar factor multiplying the state vector.

Clearly, if a stochastic state transformation with yield $p$
is possible from $\Psi$ to $\Phi$ then $\Psi \asredchem p\Phi$
because of the law of large numbers and the central limit theorem.

We are now in  a position to define the most important tool in
quantifying entanglement, namely asymptotic equivalence.
We say that $\Psi^{\otimes x}$ and $\Phi^{\otimes y}$, with $x,y\ge0$,
 are {\em asymptotically equivalent} ($\Psi^{\otimes
x}\approx\Phi^{\otimes y}$) if and only if  $\Phi^{\otimes y}$ is
asymptotically reducible to $\Psi^{\otimes x}$ and vice versa.
Two states are said to be {\em asymptotically incomparable\/} if neither is
asymptotically reducible to the other. 

Although we will mainly be concerned with asymptotic equivalence 
($\approx$), two possibly stronger
reducibilities mentioned earlier---asymptotic LOCC reducibility
with a catalyst ($\asred\dnn{LOCCc}$) and asymptotic LOCC reducibility
with a small amount of quantum communication ($\asred\dnn{LOCCq}$)---give
rise to their own corresponding versions of equivalence and incomparability. 
Since $\asred\dnn{LOCCc}$ transformations can simulate both 
$\asred\dnn{LOCCq}$ and $\asred\dnn{LOCC}$, the $\approx\dnn{LOCCc}$ 
reducibility can be expected to give rise to the simplest
(coarsest) classification of states into equivalence classes, and the
simplest (fewest independent components) entanglement
measures for multipartite states.  It has very recently been 
shown~ \cite{lin:pop:sch:wes:99} that even
$\asred\dnn{LOCCc}$ is not coarse enough to connect every isentropic
pair of states.  (The converse---that asymptotically LOCCc-equivalent
states must be isentropic---follows from the nonincrease of pure states' partial
entropies under LOCC: if $\Psi$ can be efficiently converted into
$\Phi$, even asymptotically and even with the help of a catalyst, then for each subset $X$ of the parties, $S_X(\Phi)$
cannot exceed $S_X(\Psi)$; otherwise an increase of partial entropy
could be made to occur in violation of Lemma \ref{lem:noninc} .)

\vbox{
\begin{figure}
\epsfxsize=7.8cm
\epsfbox{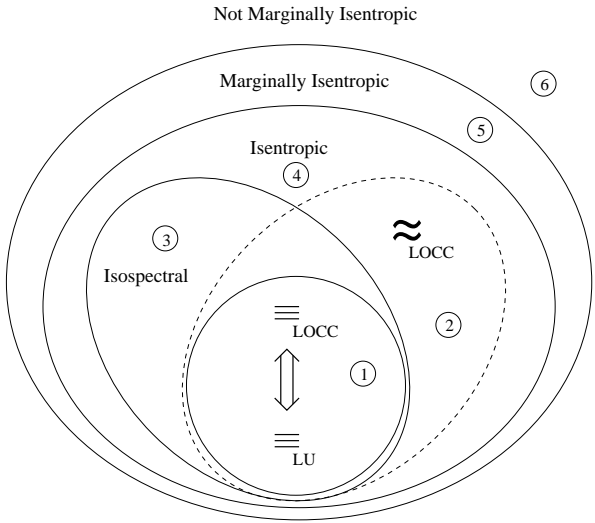}
\smallskip
\caption{Relation of exact and asymptotic equivalences to equality of
local entropies. Two states are exactly equivalent under local
operations and classical communication (LOCC) if and only if they are
equivalent under local unitary operations alone (LU). An example
(circled 1) is the LU interconvertibility of the two Bell states
$\ket{00}+\ket{11}$ and $\ket{00}-\ket{11}$. Exact equivalence of course
implies asymptotic equivalence (dotted region) including (circled 2) the
asymptotic equivalence between an EPR pair and an isentropic but not
isospectral two-trit state of the form
$\alpha\ket{00}+\beta\ket{11}+\gamma\ket{22}$. Asympotitically
equivalent states are necessarily isentropic, but not conversely. For
example (circled 3) the isentropic---and indeed isospectral---tripartite
states 2GHZ and 3EPR (see section \ref{sec:EPRGHZ}) have very recently
been shown~{\protect\cite{lin:pop:sch:wes:99}} to be incomparable with
respect asymptotic LOCC reducibility. This example also illustrates the
fact (cf {\protect\cite{kempe}}) that isospectral states of three or
more parties need not be LU-equivalent. A tensor product of circled 2
with circled 3 type states yields isentropic states (circled 4) that 
are neither asymptotically LOCC equivalent nor isospectral.
States that are marginally but not fully isentropic (circled 5) must be
incomparable with respect to exact LOCC reducibility. Finally, at the
periphery (circled 6) are states that are not even marginally
isentropic. These include incomparable pairs such as AB-EPR vs BC-EPR,
and properly reducible pairs such as GHZ vs EPR, but no cases of exact
or even asymptotic equivalence. }
\label{venn}
\end{figure}
}

We collect the relations we have proved in this section from
the definitions of the various reducibilities, using
Lemma \ref{lem:noninc} and Theorem \ref{theo:isenLU}, and
express them as 

\begin{theo}{\rm {\bf :}
\label{theo:redrels}
The following implications hold among the reducibilities, equivalences, and
partial entropies of a pair of multipartite pure states:
}\end{theo}
For reducibilities and entropy inequalities 
(omitting mention of the states $\Psi$ and $\Phi$ where
it will create no confusion) we have

\beq
\begin{array}{l}
(\Phi \equiv\dnn{LU}\!\Psi)\;\Rightarrow\;\leq\dnn{LOCC}\;\Rightarrow\;\asred\dnn{LOCC}\; \Rightarrow\; 
\asred\dnn{LOCCq} \;\Rightarrow\;  \asred\dnn{LOCCc} \;\Rightarrow\;\\ 
\forall_X \; S_X(\Phi)\leq S_X(\Psi).
\end{array}
\eeq

\medskip\noindent
For equivalences and entropy equalities we have

\beq
\begin{array}{l}
(\Phi \equiv\dnn{LU}\!\Psi)\;\Leftrightarrow\;\equiv\dnn{LOCC}\;\Rightarrow\;\approx\dnn{LOCC}\; \Rightarrow\; 
\approx\dnn{LOCCq} \;\Rightarrow\;  \approx\dnn{LOCCc} \;\Rightarrow\\ 
\forall_X \;S_X(\Phi)=S_X(\Psi)\;{\rm (i.e.}\;\Phi\;{\rm and}\;\Psi\;{\rm are~Isentropic)} \;\Rightarrow\\
\Phi\;{\rm and}\;\Psi\;{\rm are~Marginally~Isentropic} \;\Rightarrow \\
(\Phi \equiv\dnn{LU}\!\Psi)\; {\rm or} \; \Phi\;{\rm and}\;\Psi\;
{\rm are~LOCC~incomparable.} \\
\end{array}
\eeq
Figure 1 illustrates several of these relations.

\subsection{Bipartite entanglement: a reinterpretation}

As an example of the usefulness of these concepts let us reexpress the
bipartite pure-state entanglement result \cite{ben:ber:pop:sch:96}
in terms of asymptotic equivalence. In
this new language, any bipartite pure state $\Psi\upp{AB}$ is
asymptotically equivalent to $S_A(\Psi\upp{AB})$ EPR pairs: this is the
number of EPR pairs that, asymptotically, can be obtained from and are
required to prepare $\Psi\upp{AB}$ by classically coordinated local
operations.

In proving this result, the concepts of entanglement concentration and
dilution \cite{ben:ber:pop:sch:96} are central. The process of
asymptotically reducing a given bipartite pure state to
EPR singlet form is {\em entanglement dilution} and that of reducing
EPR singlets to an arbitrary bipartite pure state is
{\em entanglement concentration}.
Then the above result means that entanglement concentration and dilution
are reversible in the sense of
asymptotic equivalence, i.e., they approach unit efficiency
and fidelity in the limit of large number of copies $n$.
The crucial
requirement for these methods to work is the existence of the Schmidt
biorthogonal (normal or polar) form for bipartite pure states
\cite{hug:joz:woo:93}, that is, the fact that any bipartite pure
state  $\ket{\Psi\upp{AB}}$ can be written in a biorthogonal form:
\begin{equation}
\Psi\upp{AB}=\sum_i \lambda_i \pket{i\upp{A}}
{i\upp{B}} \enspace,
\end{equation}
where $\ket{i\upp{A}}$ and  $\ket{i\upp{B}}$ form
orthonormal bases in Alice's and Bob's
Hilbert space respectively, where by choice of phases of local bases
the coefficients $\lambda_i$ can be made real and
non-negative.


\section{Tripartite and Multipartite pure-state entanglement}
\label{sec:mulent}
In this section we use the tools we developed earlier to
propose a framework for quantifying multipartite pure-state
entanglement. Discussions in the last section were valid for pure as well
as mixed states. However from now on we will restrict out attention to pure
states.

In section \ref{sec:schdec} we consider the natural generalization of
the bipartite states, namely the $m$-party states with an $m$-way
Schmidt decomposition which we call $m$-orthogonal states. We show that
for each $m$ such states can be characterized by a scalar entanglement
measure, which may be iterpreted as the number of $m$-Cat states
asymptotically equivalent to the state in question.single parameter. In
section \ref{sec:measure} we introduce the concepts of entanglement
span, entanglement coefficients and minimal entanglement generating sets
(MREGS), as elements of a general framework for quantifying multipartite
pure-state entanglement. In section \ref{sec:lowbou} we derive lower
bounds on the cardinality of MREGS. In section \ref{sec:EPRGHZ} where we
study the question of interconversion between $m$-Cat and EPR states. In
section \ref{sec:unique} we show uniqueness of the entanglement
coefficients for natural MREGS possibilities for tripartite states.

\subsection{Schmidt-decomposable or $m$-orthogonal states}
\label{sec:schdec}
We consider Alice, Bob, Claire, ..., Matt as $m$ observers who
have one subsystem each of a $m$-part system in a joint
$m$-partite pure state. Some $m$-partite pure states, but not all, can be
written in a {\em $m$-orthogonal form} analogous to the Schmidt biorthogonal
form.  We call such states $m$-orthogonal or Schmidt decomposable.
Thus an $m$-partite pure state  $\ket{\Psi\upp{ABC...}}$ is
{\em Schmidt decomposable} or {\em $m$-orthogonal}
if and only if it can be written in a form
\begin{equation} \ket{\Psi\upp{ABC...M}}=\sum_i
\lambda_i \ket{i\upp{A}}\otimes{\ket{i\upp{B}}\otimes\ket{i\upp{C}}\ldots}
\otimes\ket{i\upp{M}} \enspace,
\end{equation} where
$\ket{i\upp{A}}$, $\ket{i\upp{B}}$, $\ket{i\upp{C}}$, ...,
$\ket{i\upp{M}}$ are orthonormal bases for the corresponding party. Notice
that by change of phases of local bases, each of the Schmidt
coefficients $\lambda_i$ can be made real and non-negative.
In any $m$-orthogonal  state, the reduced entropy seen by any
observer, indeed by any nontrivial subset of observers, is the same,
being given by the Shannon entropy of the squares of the Schmidt
coefficients. Already this makes it obvious that not all tripartite and higher
states are Schmidt decomposable since, for any $m>2$ it is clear
that there are pure $m$-partite states having unequal partial entropies for
the different observers.  Peres~\cite{per:95} gives necessary and sufficient
conditions for a multipartite pure state to be Schmidt decomposable.
Thapliyal~\cite{Thap98} recently gave another characterization, showing that an $m$-partite
pure state is Schmidt-decomposable if and only if each of the $m\!-\!1$ partite mixed
states obtained by tracing out one party is separable.

For such Schmidt decomposable  states,
the notions of entanglement concentration and dilution, developed
for bipartite states, generalize in a straightforward manner, so
that for an $m$-partite state $\Psi\upp{ABC...}$, the local entropy
as seen by any party, or indeed any nontrivial subset of the
parties, gives the asymptotic number of $m$-partite cat states
into which it can be asymptotically interconverted. That is, if
$\Psi\upp{ABC...M}$ is a Schmidt decomposable multipartite state,
then
\begin{equation}
\ket{\Psi\upp{ABC...M}} \asequ {\ket{{\rm Cat}\upp{ABC...M}}}^{\otimes 
S_A(\Psi\upp{ABC...M})} \enspace.
\end{equation}

Entanglement concentration on an $m$-orthogonal state
$\Psi\upp{ABC...M}$, like its bipartite counterpart, can be done by
parallel local actions of the observers, without any communication.
Starting with a number $n$ of copies of the state to be concentrated,
each party makes an incomplete von Neumann measurement, collapsing the
system onto a uniform superposition over an eigenspace of one eigenvalue
in the product Schmidt basis. After enough such states have been accumulated to
span a Hilbert space of dimension slightly more than some power $k$ of
$2^m$, another measurement suffices, with high probability, to collapse
the state onto a maximally entangled $m$-partite state in a Hilbert
space of dimension $2^{mk}$, which can then be transformed by local
operations into a tensor product of $k$ $m$-partite Cat states.

Entanglement dilution (cf. Fig.~2) proceeds in the same way as for bipartite states,
except that Alice locally prepares a supply of bipartite pure states
$\Phi\upp{A,A'}$ having the same Schmidt spectrum as the multipartite
Schmidt-decomposable state $\Psi\upp{ABC...M}$ which she wishes to
share with the other parties. Here the superscript $A,\!A'$ signifies that
both parts of this state are in Alice's laboratory, whereas her goal is to end
up with states shared among all the parties. As in bipartite entanglement
dilution, Alice then Schumacher-compresses the $A'$ part of a tensor
product of $n$ copies of $\Phi\upp{AA'}$, resulting in approximately $k$
compressed qubits, where $k/n$ asymptotically approaches $S_A(\Phi)=S_A(\Psi)$, 
the local entropy of the Schmidt-decomposable state she wishes to share. She
then teleports these $k$ compressed qubits to the other parties---Bob,
Claire, etc. The teleportation is performed not with $k$ EPR pairs, as
in ordinary teleportation, but with $k$ $m$-partite Cat states, which
she has shared beforehand with the other parties. For each of the
compressed qubits, Alice performs a Bell measurement on that qubit
and one leg of an $m$-partite Cat state, and broadcasts the two-bit
classical result to all the other parties, who then each apply the
corresponding Pauli rotation to their leg of the shared Cat state.
Finally all the other parties besides Alice apply Schumacher
decompression to their legs of the rotated Cat states, leaving the $m$
parties in a high-fidelity approximation to the $m$-partite state
$(\Psi\upp{ABC...M})^{\otimes n}$ which they wished to share. 
   
This entanglement dilution protocol requires $2 k/n$ bits of classical 
information per copy  (of the target state) to be  communicated from Alice to
 the other two parties.  Lo and Popescu in  \cite{lo:pop:99} show a bipartite 
entanglement dilution  protocol which requires $O(1/\sqrt{n})$ bits of
 communication per copy, thus  asymptotically, the classical 
communication cost per copy goes to zero for  their protocol. The question
 then is whether a similar protocol can be found for the dilution of 
$m$-Cat states into $m$-orthogonal states.  It is easy to see that 
replacing teleportation through EPR states with  teleportation through 
the $m$-partite Cat states in their protocol gives us a  protocol 
for entanglement dilution of the $m$-Cat states into $m$-orthogonal
states. This protocol again uses only $O(1/\sqrt{n})$
classical communication per copy, an asymptotically
vanishing amount.

\subsection{Framework for quantifying entanglement of
multipartite pure states}
\label{sec:measure}

Now we apply concepts of reducibilities and equivalences  \ref{sec:redequ}
in attempting to quantifying entanglement.  For general $m$-partite states,
there will be several inequivalent kinds of entanglement under asymptotically
reversible LOCC (or LOCCq or LOCCc) transformations---at least as many the number
of independently variable partial entropies for such states---and perhaps more.
However, a good entanglement measure ought to be defined so as to assign equal
entanglement (in the case of a multicomponent measure, equal in all components)
to asymptotically equivalent states.  This forms the basis of
our framework for quantifying entanglement.
\vbox{
\begin{figure}
\epsfxsize=8.2cm
\epsfbox{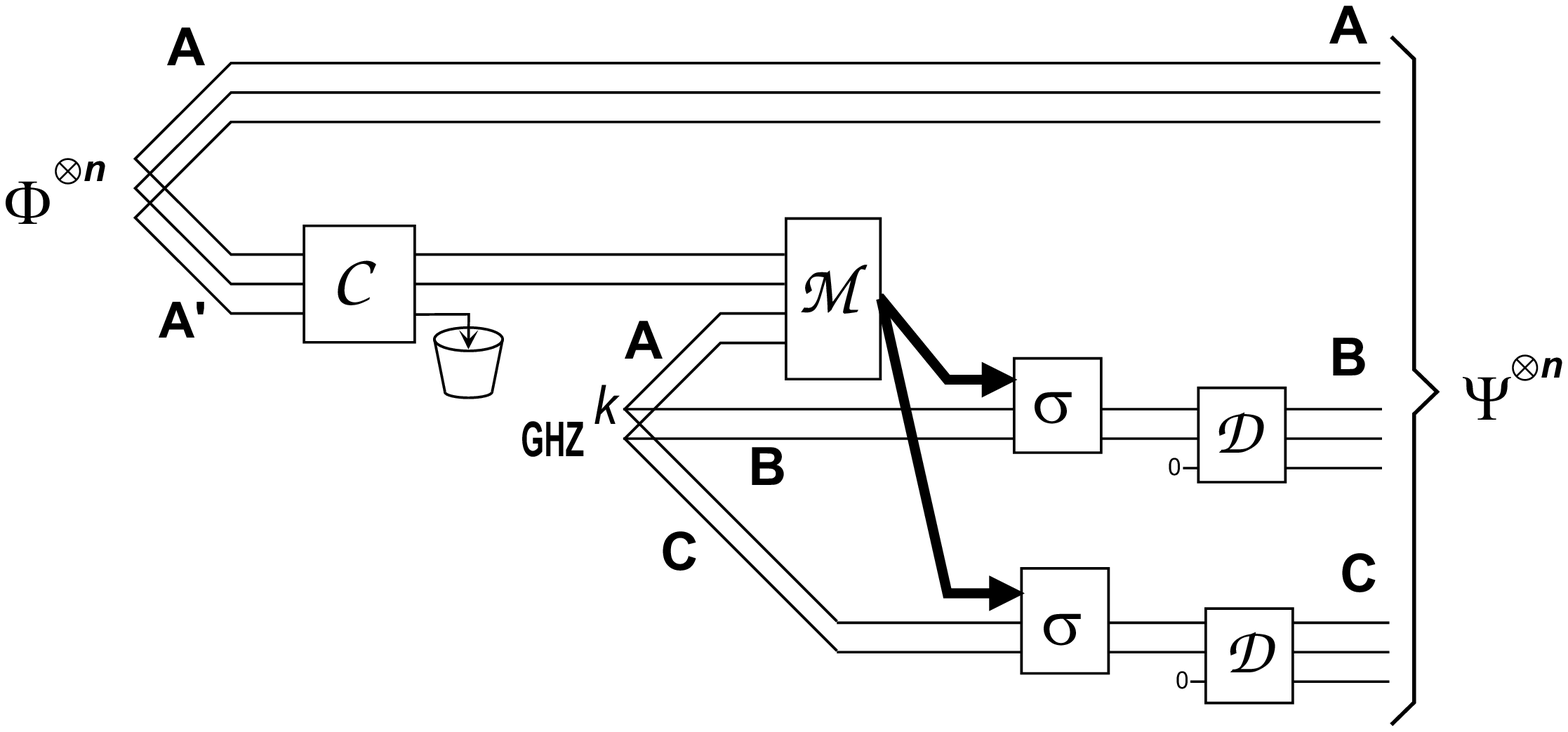}
\smallskip
\caption{Entanglement dilution for Schmidt-decomposable tripartite
states. Alice prepares a local supply of $n$ bipartite states
$\Phi\upp{AA'}$ isospectral to the Schmidt-decomposable tripartite
state $\Psi\upp{A,B,C}$ she wishes to share, and Schumacher compresses
their $A'$ halves (${\cal C}$) to $k\approx n S(\rho\dnn{A})$ qubits.
Then, using $k$ previously shared GHZ states, she teleports the
compressed qubits to Bob and Charlie simultaneously (Here ${\cal M}$
denotes a Bell measurement, the thick lines a $2k$-bit classical message
Alice broadcasts to both Bob and Claire, and $\sigma$ the conditional
Pauli rotation which completes the teleportation process). Finally, Bob
and Claire Schumacher-decompress (${\cal D}$) their $k$ qubits to
recover $n$ qubits each, in a state closely approximating $n$ copies of
the diluted Schmidt-decomposable tripartite state $\Psi\upp{ABC}$ they
wished to share.}
\label{entdil}
\end{figure}
}

We start by looking at the concept of the {\em entanglement span}
of a set of states.

Given the set of states
${\cal G}=\{\psi_1,\psi_2,...,\psi_k\}$,
their  {\em entanglement span}  ($\espan({\cal G})$) is defined as the
set of states that they can reversibly generate under asymptotic LOCC.
That is,
\begin{equation}
\espan({\cal G}) = \{\Psi \ | \ \Psi\asequ
 \bigotimes_{i=1}^{k} \ket{\psi_i}^{\otimes x_i},
\ {\mathrm with} \ x_i \ge 0\} \enspace.
\end{equation}
Notice that the $x_i$ give a quantitative amount of entanglement in
terms of the spanning states. They are called the {\em entanglement
coefficients}. In general these coefficients may be non-unique, for example
if two states in the set are locally unitarily related. Loosely speaking
these coefficients may be non-unique if the ``kinds of entanglement''
they correspond to are not ``independent''.

Let us look at some examples. The entanglement span under LOCC of
any bipartite state is the set of all bipartite states. Another example is
provided by the set of $m$-orthogonal states.
Any such state in general and in particular the $m$-Cat state
spans the set of all the $m$-orthogonal states.

Let us now introduce the concept of reversible entanglement
generating sets (REGS), which is dual to the concept of
entanglement span.
A set ${\cal G} = \{\psi_1,\psi_2,...,\psi_n$\}
of states is said to be  a {\em reversible
entanglement generating set (REGS)\/} for a class of states ${\cal C}$
if and only if ${\cal C} \subseteq \espan({\cal G})$.

Clearly, every REGS for the class of $m\!+\!1$ partite states is a REGS
for each of its $m$-partite subsystems. In particular any REGS for the full
class of $m$-partite states must be
capable of generating an EPR pair between any two of the parties.
One might suspect
that the set of all $m(m\!-\!1)/2$ EPR pairs would be a sufficient REGS for
generating all $m$-partite states, but as we will see in section
\ref{sec:lowbou}, that is not
the case for $m\!\geq\!4$.

To quantify entanglement, one would like to know the fewest kinds of
 entanglement needed to make all states in a given class.  To this end
 we define a {\em minimal reversible entanglement  generating set}
(MREGS) as a REGS of minimal cardinality.
Again the set ${\cal G}_2 =\{\rm EPR\}$ is an
example of a MREGS for bipartite entanglement which induces the
entanglement measure given by the partial entropy in bits.

Thus we have reduced the problem of quantifying entanglement to the
problem of finding the MREGS and the corresponding entanglement coefficients.
The entanglement coefficients give us the entanglement measure in terms
of how many of the states in the MREGS are required to reversibly make
the state by asymptotic LOCC. 

If we drop the requirement of reversibility, we get the notion of
a {\em entanglement generating set} (EGS), a set of
states which can generate every state in ${\cal C}$ under
exact or asymptotic LOCC.  An EGS needs only one member,
since the $m$-partite Cat state by itself is sufficient to
generate all $m$-partite entangled states, though not
reversibly.  This can be seen because the $m$-Cat state
can give an EPR pair between any two parties by exact LOCC.
So Alice can make the desired multipartite state in her lab
and then teleport it using these EPR pairs, thus generating an
arbitrary multipartite state exactly by LOCC, starting from the 
appropriate number of  $m$-Cat states.  To see that the
transformation is irreversible, note that an $m$-partite Cat state 
can be used to prepare at
most one EPR state, say between Alice and Bob, but $m\!-\!1$ EPR
states, say connecting Alice to every other party, are needed to
prepare the Cat state again.  Thapliyal~\cite{ashish1} 
has shown that a pure $m$-partite state $\Psi$ is an EGS (can be
transformed into a cat state by LOCC) if and 
only if its partial entropies $S_X$
are positive across all nontrivial partitions $X$.

The following section exhibits
some simple lower bounds on the cardinality of the MREGS for
tripartite and higher entangled pure states. Unfortunately we do
not know any corresponding upper bounds. We cannot exclude the
possibility that for tripartite and higher states an infinite
number of asympotitically inequivalent kinds of entanglement might
exist.

\subsection{Lower Bounds on the size of MREGS based on local entropies}
\label{sec:lowbou}
It is easy to see that the Alice-Bob EPR state EPR$\upp{AB}$
(regarded as a special case of an $m$-partite state in which all the parties besides
Alice and Bob are unentangled bystanders in a standard $\ket{0}$ state)
is an MREGS for the class containing all and only those states
which have $AB$ entanglement but no other entanglement, more precisely states for which
$S_X$ is zero if $X$ includes both $A$ and $B$ or neither $A$ and $B$, and has a constant
nonzero value for all other $X$. Therefore, in order to generate all possible
bipartite EPR pairs, the MREGS for general $m$-partite
pure states must have at least $m(m\!-\!1)/2$ members, which can be taken without
loss of generality to be the $m(m\!-\!1)/2$ bipartite EPR states themselves.

However, for all $m>3$ the partial entropy argument requires the MREGS
to include other states as well. Without pursuing it exhaustively~\cite{lin:pop:sch:tha:99}, 
we will sketch how local entropy arguments can be used to derive other
lower bounds on the size of the MREGS for general $m$-partite states.

Let us restrict our attention to $m$-partite pure states $\Upsilon$ in
which the partial entropy $S(\tr{X}{\proj{\Upsilon}})$ of a subset $X$
depends only on the number of members of $X$, not on which parties are
members of $X$. Two examples of such as state are the $m$-way Cat state,
and a tensor product of $m(m\!-\!1)/2$ EPR pairs, one shared between
each pair of parties. We shall call the latter an EPRs state. Let
$r_{21}(\Upsilon) = S\dnn{AB}(\Upsilon)/S\dnn{A}(\Upsilon)$ be the ratio
of two-party to one-party partial entropy in state $\Upsilon$. It is
easy to see that $r_{21}=1$ for Cat states, independent of $m$, but
$r_{21}=2(m\!-\!2)/(m\!-\!1)$ for EPRs states, the numerator of the
latter expression being the number of edges, in an $m$-partite complete
graph, joining a two-vertex subset $X$ to its complement, while the
denominator is the number of edges incident on any single vertex. Thus
Cat and EPRs states have equal $r_{21}$ for $m\!=\!3$, but for EPRs
states with larger $m$, the ratio exceeds 1, as shown in the table below.
Therefore the 4-Cat, unlike the 4-EPRs state, cannot be
asymptotically equivalent to any combination of the six EPR pairs, and
the MREGS for $m\!=\!4$ must have at least seven members.

\vbox{
\begin{table}
\begin{center}
\begin{tabular}{c|l|c}

Parties&  State                  &$r_{21}$   \\ \hline 3& Cat
(GHZ)             &1                       \\ & 3 EPRs
& 1                       \\ \hline 4&Cat                   & 1
\\ &6 EPRs          & 4/3                     \\ \hline
5&Cat                   & 1                        \\ &10
EPRs                & 3/2                     \\ &5-Qubit
Codeword               & 2                         \\ \hline 6&Cat
& 1                        \\ &15 EPRs      &  8/5
\\ \hline
\end{tabular}
\end{center}
\caption{Entropy ratio $r_{21}$ for some multipartite entangled pure states.}
\end{table}
}

\medskip\noindent
For $m=5$, the table also includes an entry for the maximally-entangled
state of five qubits, (e.g. a codeword in the well-known 5-qubit
error-correcting code~\cite{LMPZ96,ben:div:smo:woo:96}) which has maximal entropy across
any partition $X$.
Since this state has an $r_{21}$ even greater than the EPRs state, the MREGS
for $m\!=\!5$ must have at least 12 states.  Similarly, the MREGS for
$m\!=\!6$ must have at least 31 members, without considering other entropy ratios
besides $r_{21}$ or other states besides the EPR, 4-Cat, and 6-Cat states.

\subsection {Exact Reducibilities between GHZ and EPR}
\label{sec:EPRGHZ}

At this point it is natural to ask whether  three EPR pairs
(shared symmetrically among Alice, Bob, and Claire) can be
reversibly interconverted to two GHZ states. Partial entropy
arguments do not resolve the question because, for both the 3EPR
state and the 2GHZ state, the partial entropy of any nontrivial
subset of the parties is 2 bits. Nevertheless, the impossibility
of performing this conversion follows from the fact that two states 
are LOCC equivalent if and only if they are equivalent under local 
unitary operations.

To see that 2GHZ and 3EPR states are LOCC incomparable, first observe
that, since the two states are isentropic, they must, by Theorem
\ref{theo:isenLU}, either be LOCC incomparable or LU equivalent. To see
that they are not LU equivalent, observe that the mixed state obtained
by tracing out Alice from the 2GHZ state, namely $\rho_{BC}(2GHZ)$, a
maximally mixed, separable state of the two parties Bob and Claire,
while the corresponding mixed state obtained from the 3EPR states,
$\rho_{BC}(3EPR)$ is a distillable entangled state, consisting of the
tensor product of an intact BC EPR pair with another random qubit held
by each party. But if 3EPR and 2GHZ were LU equivalent, Bob and Claire,
by performing their own local unitary transformations without reference
to Alice, could make $\rho\upp{BC}(3EPR)$ from $\rho\upp{BC}(2GHZ)$.
Since they cannot do this (otherwise they would be generating
entanglement by LOCC), 3EPR and 2GHZ states cannot be LU equivalent;
therefore, by corollary \ref{cor:LULOCC} they must be LOCC incomparable.

\vbox{
\begin{figure}
\epsfxsize=7.5cm
\epsfbox{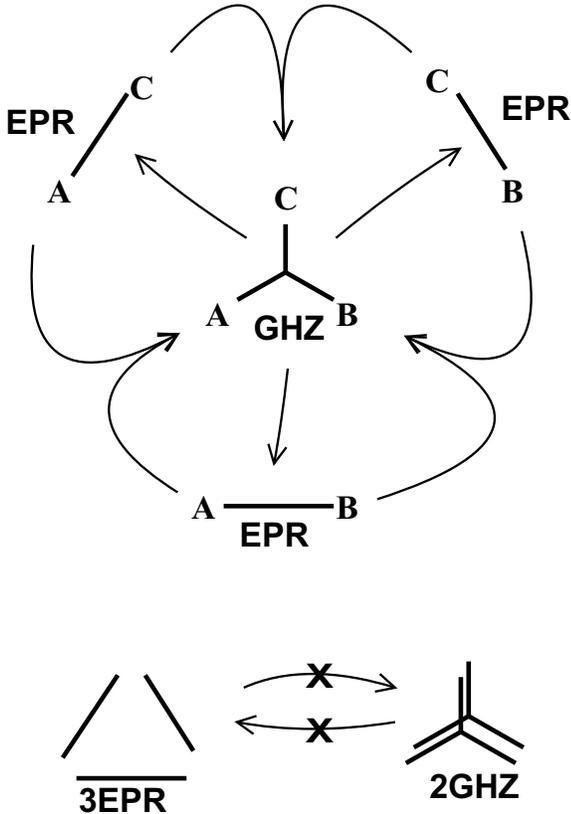}
\smallskip
\caption{Top: Two EPR pairs, together involving three parties,
can be exactly transformed to one GHZ state.
A GHZ state can be transformed into any one of the three EPR pairs.
These transformations are exact and irreversible, involving loss of entanglement
across some bipartite boundary.  Bottom: The transformations between
the symmetric 3EPR state and the 2GHZ state, marked by an X, cannot be done 
exactly, even though the partial entropies agree, by the arguments of this 
section. Very recently it has been shown {\protect \cite{lin:pop:sch:wes:99}} 
that these transformation cannot even be done asymptotically. }
\label{exreds}
\end{figure}
}

Figure 3 shows the exact reducibilities that hold among EPR and GHZ states.
The protocols for these reducibilities follow:
To get an EPR pair say between Bob and Claire, Alice performs a
measurement in the Hadamard basis namely, $\{\ket{0+1},\ket{0-1}\}$ and
informs Bob and Claire about the outcome. Using this information, Bob and
Claire can perform conditioned rotations that give them and EPR
pair. Clearly, this LOCC protocol can be generalized to many parties,
to transform a $m$-Cat state into an EPR pair between any two
parties, by having the remaining $m-2$ parties measure in the
Hadamard basis, and communicate the result to the two parties, who
then perform appropriate conditioned local unitary operations.

To get a GHZ state from two EPR pairs say $\ket{\rm EPR \upp{AB}}$
and $\ket{\rm EPR\upp{AB}}$, Alice makes a GHZ state in her lab
and then uses the EPR pairs to teleport Bob's and Charlie's parts
to them. Clearly, this protocol can be generalized to make a
$m$-Cat state from a set of $m$ EPR pairs, shared by one party with
all the rest.

In passing we note that any set of EPR pairs that describe a
connected graph, the nodes representing parties and the edges
representing the shared EPR pairs, is an EGS. This is easy to
prove using teleportation as done above.

\subsection{Uniqueness of entanglement coefficients}
\label{sec:unique}

One key question about this framework for quantifying entanglement is
whether entanglement coefficients are unique. Surely this is to be
desired if we are to interpret the values of the coefficients
as representing the amounts of different kinds of entanglement present
in the given state.

We do not know how to show uniqueness in general, but we can show this
for some cases of interest. 

For concreteness let us consider the case of three parties, say Alice,
Bob and Claire. We noted earlier that all EPR pairs shared between
two parties must be in the MREGS, so  EPR$\upp{AB}$, EPR$\upp{BC}$ and
EPR$\upp{CA}$ must be in the MREGS.
 Let us consider the entanglement span of these three EPR pairs.
Assume that there exists a state $\Psi$ in this span such that the
entanglement  coefficients are not unique, say $(x,y,z)$ and $(a,b,c)$,
where $x$, $y$ and $z$ (resp. $a$,$b$, and $c$) denote the amounts of 
EPR$\upp{AB}$,  EPR$\upp{BC}$
and EPR$\upp{CA}$ in the two decompositions.
Then using the fact that asymptotically LOCC equivalent states must
be isentropic, we have,
\begin{eqnarray}
x+y=a+b, \ \ & \ y+z=b+c,\ \  &  \ z+x=c+a \enspace.
\end{eqnarray}
This implies that $(x,y,z)=(a,b,c)$ and thus proves uniqueness.
Clearly such an argument works for the entanglement span of
EPR pairs of more parties, because there are at most $m(m-1)/2$ EPR pairs
shared by different parties and the isentropic condition gives the same
number of independent constraints.

Now we look at the entanglement span of the above three EPR pairs and the
 GHZ state. If we assume the GHZ  belongs to the span of the EPRs then
uniqueness has already been proved. Thus let us assume that the GHZ is
asymptotically not equivalent to the EPRs. Let the non-unique entanglement
coefficients be
$(x,y,z,w)$ and $(x-\delta_x,y-\delta_y,z-\delta_z,w+\delta_w)$,
with the first three coefficients
representing the amount of the EPRs and the last representing the amount of
GHZ. Without loss of generality we can assume $\delta_w=2\delta>0$.
Again using the fact that asymptotically LOCC equivalent states must
be isentropic we have
\begin{eqnarray}
\delta_w-\delta_x-\delta_y=
\delta_w-\delta_y-\delta_z=\delta_w-\delta_z-\delta_x&=&0 \enspace.
\end{eqnarray}
Solving these equations we find that,
\begin{eqnarray}
\label{eq:conditions}
\delta_x=\delta_y=\delta_z=\delta_w/2=\delta\enspace.
\end{eqnarray}

\noindent This implies that
\beq
{\rm EPR} \upp{AB} \otimes {\rm EPR}\upp{BC} \otimes 
{\rm EPR}\upp{CA}\asequ\dnn{LOCCc}{\rm GHZ}^{\otimes 2}.
\eeq

For more complicated sets ${\cal S}$ of states, the requirement that
entanglement coefficients be positive may lead to nonuniqueness.
Because of positivity, all extremal points of ${\cal S}$ must be in the
MREGS, and for some ${\cal S}$, the number of extremal
points may considerably exceed the dimensionality of ${\cal S}$
(For example, for $n\!\geq\!3$, each interior point of a regular 
$n$-gon can be expressed in multiple ways as a convex combination of vertices).

Note that there may be
many MREGS, for example any bipartite state is as MREGS for bipartite
entanglement. So how do we decide upon a {\em canonical MREGS}?
Possible criteria include requiring the states in the MREGS to be of as low Hilbert space 
dimension as possible, and as high in partial entropy within that Hilbert space as possible.
Thus for the bipartite case the EPR state is the canonical MREGS, up to local unitary operations.

\section{Discussion and Open problems}
For bipartite pure states, the unique asymptotic measure of entanglement is
known \cite{ben:ber:pop:sch:96,bbpssw96,pop:roh:97}.  The present paper 
identifies elements of any exact or asymptotic measure of {\em multi}partite
entanglement.  For bipartite states, entanglement is a scalar:  the measure of
entanglement of a state reduces to a single number.  For multipartite states,
entanglement is a vector, i.e. there are inequivalent classes of entanglement. 
The inequivalence leads to the concept of an MREGS and the requirement that
any $m$-partite entangled state be expressible as a linear combination of the
states in the $m$-partite MREGS.  Within a class of states with equivalent 
entanglement, we seek a scalar measure of entanglement.  Five desiderata for
a scalar measure of entanglement are listed in the Introduction, and section
\ref{sec:schdec} derives such a measure for the states we call 
$m$-orthogonal states. 
In this paper, however, we focus on inequivalent classes of entanglement,
leaving many questions unanswered.

Very recently \cite{lin:pop:sch:wes:99} Linden, Popescu, Schumacher and 
Westmoreland, using a relative entropy argument, have strengthened the
result of section \ref{sec:EPRGHZ} by showing that asymptotically reversible
transformations are insufficient to interconvert 2GHZ and 3EPR (indeed the states 
remain asymptotically incomparable even with the 
help of a catalyst).  Therefore the MREGS for $m\!=\!3$  
must contain at least four states (without loss of
generality the GHZ and the three bipartite 
EPR states).  Of course we would like to know whether these resources 
are sufficient to prepare {\em all} tripartite pure states in
an asymptotically reversible fashion.

A more fundamental problem is that although we have lower bounds on the
number of inequivalent kinds of entanglement under asymptotically
reversible LOCC transformations, we know of no nontrivial upper bounds.
As noted earlier, even for tripartite states we do not know that the
number is finite. One possible approach to this problem, which we do not
explore in detail here, would be to further generalize the notion of
state by allowing tensor factors to appear with negative as well as
nonintegral exponents. A generalized state such as $({\rm
EPR}\upp{AB})^{\otimes2}\otimes({\rm GHZ})^{\otimes-0.3}$ (in chemical
notation, 2EPR$\upp{AB}$ $-$ 0.3GHZ) would thus represent a quantum
``contract'' comprising a license, asymptotically, to consume two
Alice-Bob EPR pairs along with an obligation to produce 0.3 GHZ states.
Allowing negative entanglement coefficients would also solve the problem
of nonuniqueness of entanglemnet coefficients, allowing any state to be
described as a unique, but not necessarily positive, linear combination
of states in a smaller MREGS.

The most powerful result we could hope for from approaches of this
kind would be to show that under some appropriately strengthened
(but still natural) notion of asymptotic reducibility, all
isentropic states are asymptotically equivalent.  A less ambitious
result would be to show that for simple asymptotic reducibility,
or some strengthened version of it, all isentropic states are
either equivalent or incomparable, in analogy with the fact that
all isentropic states must be either equivalent or incomparable
under {\em exact\/} LOCC reducibility (corollary \ref{cor:LULOCC}).

\section*{Acknowledgements}
We thank David DiVincenzo, Julia Kempe, Noah Linden, Barbara Terhal, 
Armin Uhlmann, and Bill Wootters for
 helpful discussions. CHB, JAS, and AVT acknowledge support from
 the USA Army Research office, grant DAAG55-98-C-0041
and AVT also under DAAG55-98-1-0366. DR acknowledges support from
the Giladi program.


\begin{references}

\bibitem{ein:pod:ros:35}
A.~Einstein, B.~Podolsky, N.~Rosen, Phys. Rev. {\bf 47}, 777 (1935).

\bibitem{sch:35}
E.~Schr$\ddot{\rm o}$dinger, Naturwissenschaften {\bf 23}, 
807-812, 823-828, 844-849 (1935).
Translation: Proc of APS, {\bf 124}, 323 (1980).




\bibitem{ben:bra:cre:joz:per:woo:93} C.~H. Bennett, G.~Brassard,
C.~Cr\'{e}peau, R.~Jozsa, A.~Peres, and W.~K. Wootters,
Phys.\ Rev.\ Lett.\ {\bf 70}, 1895 (1993).

\bibitem{ben:bra:84} C.H. Bennett and G. Brassard, ``Quantum
Cryptography: Public Key Distribution and Coin Tossing'', Proceedings
of IEEE International Conference on Computers Systems and Signal
Processing, Bangalore India, December 1984, pp 175-179.; D.~Deutsch,
A.~Ekert, R.~Jozsa, C.~Macchiavello, S. Popescu,  and A.~Sanpera,
Phys.\ Rev.\ Lett.\ {\bf 77}, 2818 (1996), {\bf 80}, 2022 (1998),
H.-K. Lo ``Quantum Cryptology'' in {\em Introduction to Quantum Computation
and Information\/} by H.-K. Lo, S. Popescu and T. Spiller (World Scientific,
Singapore 1998 ISBN 981023399X), pp. 76-119; H. Zbinden ``Experimental Quantum Cryptography''
{\em ibid.\/} pp. 120-142. 

\bibitem{ben:wie:92}
C.~H.~Bennett, S.~J.~Wiesner, Phys. Rev. Lett. {\bf 69}, 2881 (1992).

\bibitem{ben:fuc:smo:97} C.~H. Bennett, C.~A. Fuchs, and J.~A. Smolin,
``En\-tangle\-ment-Enhanced Classical Communication on a Noisy Quantum
Channel,'' in {\sl Quantum Communication, Computing and Measurement},
edited by O.~Hirota, A.~S. Holevo, and C.~M. Caves (Plenum, New York,
1997).

\bibitem{ben:sho:smo:tha:99}C.H.~Bennett, P.W.~Shor, J.A.~Smolin, 
and A.V.~Thapliyal ``Entanglement-Assisted Classical Communication over
Noisy Quantum Channels,'' {\em Phys.Rev.Lett.\/} {\bf 83,}
3081 (1999) LANL e-print {\tt quant-ph/9904023}.

\bibitem{got:97}
P.~W.~Shor, Phys. Rev. A {\bf 52}, 2493 (1995);
 D.~Gottesman,
{\sl Stabilizer Codes and Quantum Error Correction},
Ph.~D. Thesis, California Institute of Technology, 1997, LANL e-print
{\tt quant-ph/9705052}.

\bibitem{deu:85}
D.~Deutsch, Proc. R. Soc. London. A {\bf 400}, 97 (1985).
D.~Deutsch, Proc. R. Soc. London. A {\bf 425}, 73 (1989).

\bibitem{gro:97}
L.~K.~Grover, LANL e-print {\tt quant-ph/9704012}.

\bibitem{cle:bur:97} R.~Cleve and H.~Buhrman,
``Substituting Quantum Entanglement for Communication,''
LANL e-print {\tt quant-ph/9704026}.

\bibitem{rev:phywor:98}Physics World, Vol. 11, No. 3, March 1998.

\bibitem{hel:kra:69:70:83}
K.-E.Hellwig and K.~Kraus, Comm. Math. Phys. {\bf 11}, 214 (1969);
K.-E.Hellwig and K.~Kraus, Comm. Math. Phys. {\bf 16}, 142 (1970);
K.~Kraus, {\em States, Effects, and Operations: Fundamental Notions of
Quantum Theory} (Springer, Berlin, 1983).

\bibitem{cav:98}
C.~M.~Caves,
Paper based on a talk presented at the {\em International Workshop
on Macroscopic Quantum   Tunneling and Coherence}, Naples, Italy, June 10-13,
1998; to be published in {\em Superconductivity}, LANL e-print
{\tt  quant-ph/9811082}.

\bibitem{ben:div:fuc:mor:rai:sho:smo:woo:98}
C.~H.~Bennett, D.~P.~DiVincenzo, C. A. Fuchs, T.~Mor, E.~Rains,
P.~W.~Shor, J.~A.~Smolin, W.~K.~Wootters, {\em Phys. Rev. A} {\bf 59}, 1070
(1999), LANL e-print {\tt quant-ph/9804053}.

\bibitem{rai:98}
E.~H.~Rains, ``A rigorous treatment of distillable entanglement,''
Phys.Rev. A {\bf 60} 173-178 (1999) LANL e-print {\tt quant-ph/9809078}.

\bibitem{ben:ber:pop:sch:96} C.~H.~Bennett, H.~J.~Bernstein, S.~Popescu,
B.~Schumacher,  Phys.\ Rev.\ A {\bf 53}, 2046 (1996),
LANL e-print {\tt quant-ph/9511030}.

\bibitem{bbpssw96} C. H. Bennett, G. Brassard, S. Popescu, B. Schumacher,
J. A. Smolin,
and W. K. Wootters, Phys. Rev. Lett. {\bf 76}, 722 (1996).

\bibitem{pop:roh:97} S. Popescu and D. Rohrlich, Phys. Rev. A 
{\bf 56}, 3319 (1997).

\bibitem{ved:ple:rip:kni:97}
V.~Vedral, M.~B.~Plenio, M.~A.~Rippin, P.~L.~Knight, Phys. Rev. Lett. {\bf 78}
2275, (1997).

\bibitem{vid:99} G. Vidal ``Entanglement Monotones," 
{\em J. Mod. Opt.\/} {\bf 47}  355 (2000),
LANL e-print {\tt quant-ph/9807077}.

\bibitem{kit:98}
A. Kitaev, private communication (1999).

\bibitem {ben:div:smo:woo:96}  C.\  H.\  Bennett, D.\  P.\  DiVincenzo,
J.\ A.\  Smolin, and W.\  K.\  Wootters, Phys.\  Rev.\ A {\bf54},
3824 (1996), LANL e-print {\tt quant-ph/9604024}.

\bibitem{lo:pop:97} H.-K. Lo and S. Popescu, ``Concentrating entanglement by
local operations---beyond mean values'', LANL e-print {\tt quant-ph/9707038}.

\bibitem{nie:98}

M.A. Nielsen, ``Conditions for a class of entanglement transformations'',
{\em Phys. Rev. Lett.\/} {\bf 83(2)}, 436--439 (1999)
LANL e-print {\tt quant-ph/9811053}.

\bibitem{jon:ple:99} 
D. Jonathan and M. B. Plenio, ``Entanglement-assisted local manipulation of
pure quantum states,'' 
{\em Phys.Rev.Lett.\/} {\bf 83,} 3566 (1999)
LANL e-print {\tt quant-ph/9905071}.


\bibitem{lin:pop:97}
N.~Linden, S.~Popescu, 
{\em Fortsch. Phys.\/} {\bf 46}, 567-578  (1998) 
LANL e-print {\tt quant-ph/9711016};
N.~Linden, S.~Popescu, A.~Sudbery, 
{\em Phys. Rev. Lett.\/} {\bf  83},  243-247 (1999)
LANL e-print {\tt quant-ph/9801076}.

\bibitem{ben:97}
C.~H.~Bennett,
``Classical and Quantum Information Transmission and Interactions'',
pp. 25-40 in {\em Quantum Communication, Computing, and Measurement}
(Proceedings of the Third International Conference on Quantum Communication
and Measurement, Sept. 1996, Shizuoka, Japan), edited by O.~Hirota,
A.~S.~Holevo, and C.~M.~Caves (Plenum, New York 1997), ISBN 0-306-45685-0.

\bibitem{weh:78}
A.~Wehrl, Rev. Mod. Phys. {\bf 50}, 221 (1978).

\bibitem{kempe} J. Kempe,  Phys.Rev. A {\bf 60}, 910-916 (1999);
LANL e-print {\tt quant-ph/9902036}.

\bibitem{uhl:76}
A.~Uhlmann, Rep.~Math.~Phys. {\bf 9}, 273 (1976).

\bibitem{joz:94}
R. Jozsa, J. Mod. Opt. {\bf 41}, 2315 (1994).

\bibitem{per:95} A. Peres, 
{\em Phys.Lett.\/} {\bf A202,} 16-17 (1995),
LANL e-print {\tt quant-ph/9504006}.

\bibitem{Thap98} A.V. Thapliyal, ``On Multipartite Pure-State Entanglement'',
{\em Phys.Rev. A\/} {\bf 59} 3336 (1999) 
LANL e-prineprint {\tt quant-ph/9811091}

\bibitem{ashish1} A.V. Thapliyal, in preparation (1999).

\bibitem{lin:pop:sch:tha:99} N. Linden, S. Popescu, B. Schumacher, A. Thapliyal,
``Partial Entropy Ratios and Multipartite Entanglement" (in preparation 1999), 
obtain many other results on the partial entropy classification of multipartite states.

\bibitem{hug:joz:woo:93}  L.~P.~Hughston, R.~Jozsa, W.\ K.\ Wootters,
Phys.\ Lett.\ A {\bf 183}, 14 (1993).

\bibitem{LMPZ96} R. Laflamme, C. Miquel, J. P. Paz, and W. H. Zurek,
Phys.~Rev.~Lett. {\bf 77}, 198 (1996).

\bibitem{lo:pop:99}
``The classical communication cost of entanglement manipulation: Is entanglement
an inter-convertible resource?''
H.-K. Lo, Sandu Popescu, Phys. Rev. Let. {\bf 83}, pp. 1459-1462, 
LANL eprint {\tt quant-ph/9902045}. 

\bibitem{lin:pop:sch:wes:99}
N. Linden, S. Popescu, B. Schumacher, M. Westmoreland, 
LANL eprint { \tt quant-ph/9912039}. 




\end{references}
\end{document}